\documentclass{PoS}

\usepackage{amsmath,amssymb,fontenc,times,mathptmx, graphicx}

\title{Exact Correlators in the 't Hooft Limit of the Principal Chiral Model}

\ShortTitle{Correlators in 't Hooft Limit...}

\author{\speaker{Peter Orland}\thanks{Supported in part by NSF Award No. PHY-0855387,  and
a PSC-CUNY Research Award.}\\
        Baruch College and the Graduate School and University Center, the City University of New York, New York, NY\\
        E-mail: \email{orland@nbi.dk}}

\abstract{The properties of $(N\times N)$-matrix-valued-field theories, in the limit $N\rightarrow \infty$, are 
harder to obtain than those for 
isovector-valued field theories. This is because we know less about the sum of planar diagrams  
than the sum of bubble/linear diagrams. Combining the $1/N$-expansion with
the axioms for form factors, exact form factors can be found for the integrable field theory of an 
SU($N$)-valued field in $1+1$ dimensions. These 
form factors can be used to find the vacuum expectation value of
the product of two field operators. We briefly
mention how the results can be applied to $2+1$ dimensional gauge theories.
}

\FullConference{The 30th International Symposium on Lattice Field Theory: Lattice 2012\\
		June 24-29, 2012\\
		Cairns, Australia}

\begin{document}

\section{Introduction}

What is frustrating about analytic attempts to solve quantum field theory non-perturbatively is 
that the ``exciting new ideas" take many years to be put to use, if at all. An example is the 
large-$N$ limit of 't Hooft. 't Hooft's idea remains promising for quantum chromodynamics 
(QCD), but is four decades old. There is no solution of $N\rightarrow \infty$ theories with
dynamical mass gaps (though for $(1+1)$-dimensional 
QCD, matrix integrals, matrix quantum mechanics and
maximally supersymmetric conformal field theories, the situation is better). Despite the 
gloom, optimism still lives; hence this contribution. 

The integrable principal chiral sigma model has a field $U(x)$, lying in the space of unitary $N\times N$ matrices of determinant one, with $x=(x^{0}, x^{1})$, being a point of Minkowski space, with metric
$\eta_{00}=-\eta_{11}=1$, $\eta_{01}=\eta_{10}=0$.  The action is
\begin{eqnarray}
S=\frac{N}{2g_{0}^{2}}\int d^{2}x \;\eta^{\mu\nu}\;{\rm Tr}\,\partial_{\mu}U(x)^{\dagger}\partial_{\nu}
U(x),
\label{action}
\end{eqnarray}
where $\mu, \nu=0,1$. The action is invariant under 
$U(x)\rightarrow V_{L}U(x)V_{R}$,  for two constant $N\times N$ unitary matrices $V_{L}, \,V_{R}$, with a specified overall
phase (hence the global symmetry is ${\rm U}(N)\times {\rm U}(N)/{\rm U}(1)$). We do not 
include a Wess-Zumino-Witten term in this action. This theory is asymptotically free, and
we assume it has a mass gap $m$, for all $g_{0}$ and 
$N\ge 2$. In the limit $N\rightarrow \infty$, the Feynman diagrams for Green's functions 
become planar, just as they do for QCD. 

The S matrix of the action (\ref{action}) was found in the nineteen-eighties \cite{PWetc}. The methods used to find this
S matrix relied heavily on the integrability of the principal chiral model. The S matrix does not tell the whole story, however. In 
a theory of something-but-not-everything, we need to know off-shell 
behavior. This behavior
is encoded in the form factors of the theory \cite{smirnov}, which tell us how quantum states respond to external 
probes. Form factors may be used to determine vacuum expectation values.

Some form factors for the scaling field were found in Ref. \cite{summing1}, and
the rest in Ref. \cite{summing2}. Form factors for currents were found in Ref. \cite{axel}.

Isovector field theories are much easier to solve in the limit of a large number of components. An example is the O($N$) sigma model in $d$ space-time dimensions, whose field $r(x)$ is an
$N$-component column vector, satisfying $r^{\rm T}(x)r(x)=1$, with action
\begin{eqnarray}
S=\frac{N}{2g_{0}^{2}}\int d^{d}x \;\eta^{\mu\nu}\, \partial_{\mu}r^{\rm T}(x)\partial_{\nu} r(x).
\label{vecmodel}
\end{eqnarray}
The Feynman diagrams have a linear structure. They can be summed
at $N=\infty$, and $1/N$ corrections are straightforward. We 
know a great deal about such 
models for large $N$, even in the non-renormalizable case of more than 
two dimensions, where there is an ultraviolet-stable fixed point.

In simpler models like (\ref{vecmodel}), the Green's functions are those of massive free 
fields in the large-$N$ limit, with the 
mass gap depending on the coupling $g_{0}$ and the ultraviolet regulator. This is completely false for
matrix models like (\ref{action}), despite the fact the S matrix becomes unity in the limit. Furthermore, there 
is an infinite renormalization of the components of
$U(x)$ at $N=\infty$;  this does not happen until order $1/N$ for isovector models. There does exist a massive free ``master" field in the $N \rightarrow \infty$ limit of the principal chiral model, but its relation to $U(x)$ is nontrivial.

For up-to-date reviews of large-$N$ investigations, see Ref. \cite{panero}.

My interest in this problem came from studying confinement, viewing 
non-Abelian gauge theories
as coupled $(1+1)$-dimensional principal chiral 
sigma models. This connection can be understood 
in a number of different ways, going back many years \cite{YMsigma}. It is particularly interesting
in $2+1$ dimensions, because no strong-coupling assumption is needed \cite{durhuus-fr}. Using
exact form factors known for the principal chiral model whose field lies in SU($2$) \cite{KarowWiesz}, the
static quark-antiquark potential can be found, albeit in an anisotropic theory \cite{hor-string}, confirming
the earlier suggestions of Ref. \cite{YMsigma}. The mass spectrum and $k$-string
tensions can also be found \cite{mass-spectrum}. For this reason, it seems worthwhile to generalize
results for the principal chiral model beyond SU($2$). Remarkably, the large-$N$ principal chiral
model seems simpler than the SU($2$) case.

In the next section we summarize the results of Refs. \cite{summing1,summing2,axel}. In Section 3, we 
discuss the relevance of these results to lower-dimensional gauge theories.

\section{Form Factors, Green's Functions and the Master Field}

We first introduce the scaling field $\Phi(x)$. This
is a complex $N\times N$ matrix, obeying no specific constraints. This field is
apparent in ordinary perturbation theory, as well as in our approach. Matrix elements
of $\Phi(x)$ and $\Phi(x)^{\dagger}$ are related to those of $U(x)$ and $U(x)^{\dagger}$
through 
\begin{eqnarray}
\Phi(x)\sim Z^{-1/2}U(x),\;\Phi(x)^{\dagger}\sim Z^{-1/2}U(x)^{\dagger},
\end{eqnarray}
where $Z$ is a real infinitesimal renormalization factor. 

The two-point Wightman function is
\begin{eqnarray}
&\!\!\frac{1}{N}&\!\!\!\langle 0\vert \Phi(0) \Phi(x)^{\dagger}\vert 0\rangle=\frac{1}{NZ}\langle 0\vert U(0) U(x)^{\dagger}\vert 0\rangle
=\int \frac{d\theta_{1}}{4\pi} e^{{\rm i}m(x^{-}e^{\theta_{1}}+x^{+}e^{-\theta_{1}})}   \nonumber \\
\!\!&\!\!+\!\!&\!\! \frac{1}{4\pi}\sum_{l=1}^{\infty}\int d\theta_{1}\cdots d\theta_{2l+1}
\exp\!\!\left[ {\rm i}\sum_{j=1}^{2l+1}m (x^{-}e^{\theta_{j}}+x^{+}e^{-\theta_{j}})\!\!\right]
\!\!\prod_{j=1}^{2l}\frac{1}{(\theta_{j\;j+1})^{2}+\pi^{2}} +O(1/N),
\label{series}
\end{eqnarray}
where $x^{\pm}=(x^{0}\pm x^{1})/2$. The quantities $\theta_{1},\dots,\theta_{2l+1}$ are rapidities of excitations, related
to the momenta of these excitations by $p_{j}=(m\cosh\theta_{j},m\sinh \theta_{j})$. A rapidity variable with two subscripts denotes the difference of two rapidities, $\theta_{jk}=\theta_{j}-\theta_{k}$. The excitations themselves are particles or antiparticles, with two colors (or alternatively one color and one anti-color) which
have $N$ possible values. A free-field correlator has only the first term. The terms in this series (\ref{series}) converge, by virtue
of being the result of repeated applications of the Poisson kernel (for more discussion, see Ref.
\cite{summing2}). What makes this result possible is the absence of bound states in the large-$N$ limit. 
The particles of the principal chiral model consist of ``elementary" particles of mass $m$
and bound states of $r$ of
these particles. The masses of these bound states are given by the sine law
$m_{r}=m\sin\frac{\pi r}{N}/sin \frac{\pi}{N}$. If we assume that $m$ is fixed as $N\rightarrow \infty$, the binding
energy vanishes for all of these, except the antiparticle, for which $r=N-1$.

To find form factors, we need the two-particle S matrix \cite{PWetc}. We won't explain its origin here (a guide through the literature is
presented in Section 1 of
Ref. \cite{summing1}). This S matrix has the $1/N$ expansion
\begin{eqnarray}
S_{PP} (\theta)
=\left[ 1+O(1/N^{2})\right]
\left[1-\frac{2\pi {\rm i}}{N\theta}(P\otimes 1+1\otimes P)-\frac{4\pi^{2}}{N^{2}\theta^{2}}P\otimes P
\right]. \label{expanded-s-matrix}
\end{eqnarray}
The symbol $P$ denotes the operation of interchanging colors between particles. In the second and third terms either the left or right colors are interchanged. In the last term, both are interchanged.
We can find the scattering matrix of one particle and one antiparticle $S_{AP}(\theta)$ from
(\ref{expanded-s-matrix}), using crossing.

The expression (\ref{series}) is found from an infinite set of form factors for $\Phi$. Smirnov's axioms for the form factors of integrable field theories in $1+1$ dimensions are \cite{smirnov}:
\vspace{5pt}

\noindent
{\em Scattering Axiom} (Watson's theorem).
\begin{eqnarray}
&\!\!\!\!\langle& \!\!\!\!0\vert \Phi(0)_{b_{0}a_{0}}\vert I_{1},\theta_{1},C_{1};
\dots ; I_{j}, \theta_{j}, C_{j} ; I_{j+1}, \theta_{j+1}, C_{j+1} ;\dots ; I_{n}, \theta_{n}, C_{n} \rangle_{\rm in}
 \nonumber \\
&=&
S_{I_{j}I_{j+1}}(\theta_{j\; j+1})^{C^{\prime}_{j+1}C^{\prime}_{j}}_{\;\;\;\;\;C_{j}C_{j+1}} \nonumber \\
&\times&\langle 0\vert \Phi(0)_{b_{0}a_{0}}\vert I_{1},\theta_{1},C_{1};
\dots ; I^{\prime}_{j+1}, \theta_{j+1}, C^{\prime}_{j+1} ; I^{\prime}_{j}, \theta_{j}, C^{\prime}_{j} ;\dots ; I_{n}, \theta_{n}, C_{n} \rangle_{\rm in}
, \label{watson}
\end{eqnarray}
where $I_{k}$, $k=1,\dots,n$ is $P$ or $A$ (particle or antiparticle) and $C_{k}$ denotes a pair
of color indices (which may be written $a_{k}b_{k}$, for $C_{k}=P$ and
 $b_{k}a_{k}$, for $C_{k}=A$) and similarly for the primed indices. Though
(\ref{expanded-s-matrix}) becomes unity as $N\rightarrow \infty$, contractions of colors 
produce factors of order $N$.

\vspace{5pt}

\noindent
{\em Periodicity Axiom}.
\begin{eqnarray}
\langle \!\!\!&\!\!\!0\!\!\!&\!\!\!\vert \Phi(0)_{b_{0}a_{0}}\vert I_{1},\theta_{1}, C_{1}; \;
\dots ;\; I_{n},\theta_{n}, C_{n}
 \rangle_{\rm in}  \nonumber \\
&=&\!\!\!\langle 0\vert \Phi(0)_{b_{0}a_{0}}\vert
I_{n}, \theta_{n}-2\pi{\rm i}, C_{n} ;\; I_{1},\theta_{1}, C_{1}; \;
\dots ; \;I_{n-1},\theta_{n-1}, C_{n-1}
 \rangle_{\rm in} .
\label{periodicity}
\end{eqnarray}

\vspace{5pt}

\noindent
{\em Annihilation-Pole Axiom}.  This fixes the residues of the poles of the form 
factors. 
\begin{eqnarray}
&{\rm  Res}\!\!&\!\!\vert_{\theta_{1n}=-\pi{\rm i}} \,
\langle \, 0 \,\vert\, \Phi(0)_{b_{0}a_{0}}
\vert I_{1},\theta_{1}, C_{1}; \; I_{2},\theta_{2}, C_{2};
\dots ;\;I_{n},\theta_{n}, C_{n}
 \rangle_{\rm in}   \nonumber \\
&=&-2{\rm i} \langle \, 0 \,\vert\, \Phi(0)_{b_{0}a_{0}}
\vert  I_{2},\theta_{2}, C^{\prime}_{2}; \;I_{3},\theta_{3}, C^{\prime}_{3};\;
\dots ;\; I_{n-1},\theta_{n-1}, C^{\prime}_{n-1} \rangle_{\rm in}  
 \nonumber \\
&\times&
\left[  S_{I_{1}I_{2}}(\theta_{12})^{C_{2}^{\prime}D_{1}}_{\;\;C_{1}C_{2}}
 S_{I_{1}I_{3}}(\theta_{13})^{C_{3}^{\prime}D_{2}}_{\;\;D_{1}C_{3}}
\cdots S_{I_{1}I_{n-1}}(\theta_{1\;n-1})^{C_{n}C_{n-1}^{\prime}}_{\;\;D_{n-2}C_{n-1}}
-\delta^{C_{n}}_{\;\;C_{1}}
\delta^{C_{2}^{\prime}}_{\;\;C_{2}}\delta^{C_{3}^{\prime}}_{\;\;C_{3}}
\cdots \delta^{C_{n-1}^{\prime}}_{\;\;C_{n-1}}
\right], 
\label{a-p}
\end{eqnarray}

\vspace{5pt}

\noindent
{\em Lorentz-Invariance Axiom}. For the scalar operator $\Phi$, this takes the form
\begin{eqnarray}
\langle \!\!\!&\!\!\!0\!\!\!&\!\!\! \vert\, \Phi(0)_{b_{0}a_{0}}
\vert I_{1},\theta_{1}+\Delta\theta, C_{1}; \; I_{2},\theta_{2}+\Delta \theta, C_{2};
\dots ;\;I_{n},\theta_{n}+\Delta \theta, C_{n}
 \rangle_{\rm in} \nonumber \\
&=&
\langle \, 0 \,\vert\, \Phi(0)_{b_{0}a_{0}}
\vert I_{1},\theta_{1}, C_{1}; \; I_{2},\theta_{2}, C_{2};
\dots ;\;I_{n},\theta_{n}, C_{n}
 \rangle_{\rm in}  ,
\label{lorentz}
\end{eqnarray}
for an arbitrary boost $\Delta \theta$.

\vspace{5pt}

\noindent
{\em Bound-State Axiom}. There are poles on the imaginary axis of 
rapidity differences
$\theta_{jk}$, due to bound 
states. We can ignore this axiom as $N\rightarrow \infty$.

\vspace{5pt}

\noindent
{\em Minimality Axiom}. The
form factors have as much analyticity in the complex strip 
$0<{\mathfrak Im}\;\theta_{jk}<2\pi$ as is consistent with the other axioms.

We now describe the form factors of $\Phi(0)$, from which (\ref{series}) was obtained. These are 
consistent with the axioms discussed above. They are 
\begin{eqnarray}
\langle \!\!\!\!\!&\!\!\!\!\!0\!\!\!\!\!\!&\!\!\!\! \vert \Phi(0)_{b_{0}a_{0}}\;
\vert A,\theta_{1},b_{1},a_{1};\dots; A, \theta_{M-1}, b_{M-1}, a_{M-1};
P,\theta_{M}, a_{M},b_{M};\dots; P,\theta_{2M-1}, a_{2M-1,b_{2M-1}}
\rangle_{\rm in} 
\nonumber \\
&=&
N^{-M+1/2}\sum_{\sigma,\tau\in S_{M}}
F_{\sigma \tau}(\theta_{1},\theta_{2},\dots,\theta_{2M-1})
\prod_{j=0}^{M-1}\delta_{a_{j}\;a_{\sigma(j)+M}} 
\delta_{b_{j}\;b_{\tau(j)+M}} ,
\label{special-FF}
\end{eqnarray}
for a particular function $F$. We now try to clarify the meaning of this complicated expression. The permutations
$\sigma$ and $\tau$ are elements of the permutation group of $M$ objects $S_{M}$; they act
on permutations of $0,1,\dots,M-1$. The rapidities of the excitations are ordered by $\theta_{1}<\theta_{2}<\cdots<\theta_{2M-1}$. The letter $A$ denotes an antiparticle, and $P$ denotes a particle.The indices $a_{j}$ and $b_{j}$ are left and right
color indices, respectively. Other orderings of the rapidities change the right-hand side by an
overall phase. Replacing $\Phi(0)$ by $\Phi(x)$ is done by multiplying by 
$\exp -{\rm i} \sum_{j=1}^{2M-1}
p_{j}\cdot x_{j}$. The function $F$ is given to leading order in $1/N$ by
\begin{eqnarray}
F_{\sigma \tau}(\theta_{1},\theta_{2},\dots,\theta_{2M-1})
=
\frac{ (-4\pi)^{M-1}K_{\sigma \tau} }{\prod_{j=1}^{M-1} 
[\theta_{j}-\theta_{\sigma(j)+M}+\pi{\rm i}][\theta_{j}-\theta_{\tau(j)+M}+
\pi{\rm i}]}+O(1/N),  \label{MFF}
\end{eqnarray}
where 
$K_{\sigma \tau}$ is unity if $\sigma(j)\neq \tau(j)$, for all$j$, and zero otherwise.
All other form factors can be obtained from these expressions by crossing.

The series (\ref{series}) is obtained from the form factors (\ref{special-FF}), (\ref{MFF})
by using the completeness relation for in-states
$\vert X\rangle_{in}$:
\begin{eqnarray}
\langle 0 \vert \Phi(0) \Phi(x)^{\dagger} \vert 0\rangle
= 
\sum_{X} \langle 0 \vert \Phi(0)\vert X\rangle_{\rm in\;in}  \langle X\vert \Phi(x)^{\dagger} \vert 0\rangle
\end{eqnarray}

Form factors can be obtained for other operators besides fields. The Lorentz-invariance axiom is modified by the inclusion of spin. A. Cort\'{e}s Cubero \cite{axel} has found some the form factors
for the Noether current of the left symmetry $j^{\rm L}(x)_{\mu}={\rm i}\, \partial_{\mu}U(x)\, U(x)^{\dagger}$. The results for the current of the right symmetry
$j^{\rm R}_{\mu}(x)={\rm i} \, U(x)^{\dagger}\,\partial_{\mu}U(x)$ are similar. The results are
\begin{eqnarray}
\langle\!\!&\!\! 0 \!\!&\!\! \vert  j^{L}_{\mu}(0)_{a_{0}c_{0}}\vert
A,\theta_{1}, b_{1}, a_{1};P,\theta_{2},b_{2},a_{2}
\rangle_{\rm in} \nonumber \\
&=&(p_{1}-p_{2})_{\mu}\;\frac{2\pi {\rm i}}{\theta_{12}+\pi{\rm i}}
\left(\delta_{a_{0}a_{2}}\delta_{c_{0}a_{1}}
-\frac{1}{N}\delta_{a_{0}c_{0}}\delta_{a_{1}a_{2}}
\right)\delta_{b_{1}b_{2}}+O\left(\frac{1}{N^{2}}\right).
\nonumber
\end{eqnarray}
and
\begin{eqnarray}
\langle \!\!&\!\!0\!\!&\!\! \vert j^{L}_{\mu}(0)_{a_{0}c_{0}}
\vert A,\theta_{1}, b_{1}, a_{1};A,\theta_{2},b_{2},a_{2};
P,\theta_{3}, a_{3}, b_{3};P,\theta_{4},a_{4},b_{4}\rangle_{\rm in} \nonumber \\  
&=&\frac{8\pi^{2}{\rm i}}{N}(p_{1}+p_{2}-p_{3}-p_{4})_{\mu} \nonumber\\
&\times&\left[\frac{1}{(\theta_{14}+\pi{\rm i})(\theta_{23}+\pi{\rm i})(\theta_{24}+\pi{\rm i})}
\left( 
\delta_{a_{0}a_{3}}\delta_{a_{1}c_{0}}-\frac{1}{N}\delta_{a_{0}c_{0}}\delta_{a_{1}a_{3}}
\right)\delta_{a_{2}a_{4}}\delta_{b_{1}b_{4}}\delta_{b_{2}b_{3}}
\right.
\nonumber \\
&+&\frac{1}{(\theta_{13}+\pi{\rm i})(\theta_{23}+\pi{\rm i})(\theta_{24}+\pi{\rm i})}
\left( 
\delta_{a_{0}a_{4}}\delta_{a_{1}c_{0}}-\frac{1}{N}\delta_{a_{0}c_{0}}\delta_{a_{1}a_{4}}
\right)\delta_{a_{2}a_{3}}\delta_{b_{1}b_{3}}\delta_{b_{2}b_{4}} \nonumber \\
&+&\frac{1}{(\theta_{14}+\pi{\rm i})(\theta_{13}+\pi{\rm i})(\theta_{24}+\pi{\rm i})}
\left( 
\delta_{a_{0}a_{3}}\delta_{a_{2}c_{0}}-\frac{1}{N}\delta_{a_{0}c_{0}}\delta_{a_{2}a_{3}}
\right)\delta_{a_{1}a_{4}}\delta_{b_{1}b_{3}}\delta_{b_{2}b_{4}} \nonumber \\
&+&\left. \frac{1}{(\theta_{14}+\pi{\rm i})(\theta_{13}+\pi{\rm i})(\theta_{23}+\pi{\rm i})}
\left( 
\delta_{a_{0}a_{4}}\delta_{a_{2}c_{0}}-\frac{1}{N}\delta_{a_{0}c_{0}}\delta_{a_{2}a_{4}}
\right)\delta_{a_{1}a_{3}}\delta_{b_{1}b_{4}}\delta_{b_{2}b_{3}}\right] \nonumber \\
&+&O\left(\frac{1}{N^{3}}\right).
\end{eqnarray}
It seems that all the form factors of the current operator and the stress-energy
tensor can be obtained \cite{future}.

To describe the master field, we need some more formalism. An in-state is defined as a product of creation operators in the order of increasing rapidity, from right to left, acting on the vacuum, {\em e.g.}
\begin{eqnarray}
\vert P,\theta_{1},a_{1},b_{1};A,\theta_{2},b_{2},a_{2},\dots \rangle_{\rm in}
={\mathfrak A}^{\dagger}_{P}(\theta_{1})_{a_{1}b_{1}}
{\mathfrak A}^{\dagger}_{A}(\theta_{2})_{b_{2}a_{2}}\cdots \vert 0\rangle,\;\;{\rm where}\;
\theta_{1}>\theta_{2}>\cdots
\end{eqnarray}
These creation operators satisfy  the Zamolodchikov 
algebra: 
\begin{eqnarray}
{\mathfrak A}^{\dagger}_{I_{1}}(\theta_{1})_{C_{1}}\,
{\mathfrak A}^{\dagger}_{I_{2}}(\theta_{2})_{C_{2}}
&=&S_{I_{1}I_{2}}(\theta_{12})^{C_{2}^{\prime};C_{1}^{\prime}}_{C_{1};C_{2}}\;
{\mathfrak A}^{\dagger}_{I_{2}}(\theta_{2})_{C_{2}^{\prime}}\,
{\mathfrak A}^{\dagger}_{I_{1}}(\theta_{1})_{C_{1}^{\prime}}.
\label{creation}
\end{eqnarray}
The Yang-Baxter relation is a consistency condition for 
(\ref{creation}). The master field is
\begin{eqnarray}
M(x)=\int \frac{d\theta}{4\pi}\, \left[{\mathfrak A}_{P}(\theta)
e^{{\rm i}m x^{0}\cosh \theta-{\rm i}m 
x^{1}\sinh \theta} 
+{\mathfrak A}^{\dagger}_{A}(\theta)e^{-{\rm i}m x^{0}\cosh \theta+{\rm i}m x^{1}\sinh \theta} 
\right], \label{master}
\end{eqnarray}
where ${\mathfrak A}_{A}$ is the destruction operator of an antiparticle. It is the adjoint of the operator  
${\mathfrak A}^{\dagger}_{A}$. In the limit $N\rightarrow \infty$, $[{\mathfrak A}_{A,P}(\theta),
{\mathfrak A}^{\dagger}_{A,P}(\theta)]\rightarrow 4\pi \delta(\theta-\theta^{\prime})$, and all other commutators approach zero. Thus the  operator $M(x)$ is a massive free field\footnote{Witten's original observation was that there should exist a specific classical master-field configuration. There is little difference between this and a free field.} as $N\rightarrow \infty$. The form factors give the coefficients of an expansion of 
the renormalized field $\Phi(x)$ in terms of $M(x)$ \cite{summing1}.

\section{Gauge theories as coupled sigma models}

In this final section, we will briefly explain how our results may be used to study gauge theories in lower dimensions. We take the time coordinate $x^{0}$ and one space coordinate $x^{1}$ continuous, but
$x^{2}$. In axial gauge, $A_{1}=0$, or $U_{1}=1$. The remaining lattice gauge field is 
$U_{2}(x^{0},x^{1}, x^{2})$, and we drop the subscript $2$. The left-handed and right-handed currents may be redefined as
$j^{\rm L}_{\mu}(x)_{b}={\rm i}{\rm Tr}\,t_{b} \, \partial_{\mu}U(x)\, U(x)^{\dagger}$ and
$j^{\rm R}_{\mu}(x)_{b}={\rm i}{\rm Tr}\,t_{b} \, U(x)^{\dagger}\partial_{\mu}U(x)$, respectively, 
where $\mu=0,1$ and $t_{b}$ is a generator, with normalization ${\rm Tr}\,t_{a}t_{b}=\delta_{ab}$. The Hamiltonian is $H_{0}+H_{1}$, where
\begin{eqnarray}
H_{0}\!=\!\sum_{x^{2}}\int dx^{1} \frac{1}{2g_{0}^{2}}\{ [j^{\rm L}_{0}(x)_{b}]^{2}+[j^{\rm L}_{1}(x)_{b}]^{2}\}
\;,\label{HNLSM}
\end{eqnarray}
and
\begin{eqnarray}
H_{1}\!\!&\!\!=\!\!&\!\! \sum_{x^{2}}  \int dx^{1} \,
\frac{(g_{0}^{\prime})^{2}a^{2}}{4}\,[\partial_{1}\Phi(x^{1},x^{2})_{b}]^{2} 
+(g_{0}^{\prime})^{2}q_{b}\Phi(u^{1},u^{2})_{b} -(g_{0}^{\prime})^{2}
q^{\prime}_{b}\Phi(v^{1},v^{2})_{b} \nonumber \\
\!\!&\!\!-\!\!&\!\! 
\left(\frac{g_{0}^{\prime}}{g_{0}}\right)^{2}\,\,\sum_{x^{2}=0}^{L^{2}-a}  \int dx^{1} \!\!
\left[ j^{\rm L}_{0}(x^{1},x^{2})_{b}\Phi(x^{1},x^{2})_{b} -j^{\rm R}_{0}(x^{1},x^{2})_{b} \Phi(x^{1},x^{2}+a)_{b} \right]  
  \; ,
\label{continuum-local}
\end{eqnarray}
where $-\Phi_{b}=A_{0\,\,b}$ is the temporal gauge field, and
where in the last term
we have inserted two color charges - a quark with charge $q$ at site $u$
and an anti-quark with charge $q^{\prime}$ at site $v$. Some gauge invariance remains
after the axial-gauge fixing, namely that 
for each $x^{2}$
\begin{eqnarray}
\left\{ \int d x^{1}\left[ j^{L}_{0}(x^{1},x^{2})_{b}-j^{R}_{0}(x^{1},x^{2}-a)_{b}\right] - g_{0}^{2}Q(x^{2})_{b} \right\}\Psi=0\;,
\label{physical}
\end{eqnarray}
on wave functionals $\Psi$, 
where $Q(x^{2})_{b}$ is the total color charge from quarks at $x^{2}$ and $\Psi$ is any physical 
state. To derive the constraint (\ref{physical}) more precisely, we started with open boundary
conditions in the $1$-direction and periodic boundary conditions in
the $2$-direction, meaning that the two-dimensional space is a cylinder.

The unperturbed Hamiltonian (\ref{HNLSM}) is a discrete sum of principal-chiral nonlinear sigma model 
Hamiltonians. Ultimately, we would like $g_{0}^{\prime}=g_{0}$, and take both parameters to zero. Unfortunately making approximations using form factors rely on $g_{0}^{\prime}\ll g_{0} \ll1$
\cite{hor-string}. The critical point $g_{0}=g_{0}^{\prime}=0$ can be approached,
but along a curve, which is tangent to the $g_{0}$-axis, in a graph of $g_{0}$ 
versus $g_{0}^{\prime}$. Thus, we need to understand the crossover to the isotropic theory with $g_{0}^{\prime}\approx g_{0}$. For large $N$, this problem simplifies, because the eigenstates
of (\ref{HNLSM}) are free-particle states. Ultimately, a real-space renormalization in the $x^{2}$ direction
\cite{konik}
may be necessary to understand the isotropic theory.

\end{document}